# EXPERIMENTAL STUDIES OF STABLE CONFINED ELECTRON CLOUDS USING GABOR LENSES

O. Meusel, M. Droba, B. Glaeser, K. Schulte
IAP, University Frankfurt/Main, Germany

*Abstract*
  Based on the idea of D. Gabor [1] space charge lenses are under investigation to be a powerful focussing device for intense ion beams. A stable confined electron column is used to provide strong radially symmetric electrostatic focussing, e.g. for positively charged ion beams.
  The advantages of Gabor lenses are a mass independent focussing strength, space charge compensation of the ion beam and reduced magnetic or electric fields compared to conventional focussing devices. Collective phenomena of the electron cloud result in aberrations and emittance growth of the ion beam. The knowledge of the behaviour of the electron cloud prevents a decrease of the beam brilliance. Numerical models developed to describe the electron confinement and dynamics within a Gabor lens help to understand the interaction of the ion beam with the electron column and show the causes of non-neutral plasma instabilities.
  The diagnosis of the electron cloud properties helps to evaluate the numerical models and to investigate the influence of the ion beam on the confined non-neutral plasma.

## INTRODUCTION

The focal strength of a space charge lens is determined by the spatial density distribution of the confined non-neutral plasma. It is a function of the transverse and longitudinal trapping conditions. For a radial confinement by a magnetic field, Gabor showed that in absence of external electric fields the transverse trapping condition is given by the Brillouin flow limit and therefrom the maximum electron density can be calculated by:

$$n_{e,rad} = \frac{\varepsilon_0}{2 m_e} B_z^2 \qquad (1)$$

The upper limit for the longitudinal confinement is given by the space charge potential of a homogeneous cloud of radius $r_A$, which has to be smaller than the anode potential $V_A$, resulting in:

$$n_{e,l} = \frac{4\varepsilon_0 V_A}{e r_A} \qquad (2)$$

Each of these criteria solely overestimates the space charge density significantly. Additionally, the longitudinal confinement condition is drastically influenced by thermalisation of the enclosed particles and therefrom due to losses of fast particles in the Maxwellian tail. The measured focal length f of the Gabor lenses gives the possibility to determine the average electron density inside the lens. Applying the short lens approximation to a homogeneous space charge cloud the average electron density is given by:

$$\overline{n}_e = \frac{4\varepsilon_0 W_B}{f \cdot e} \cdot l \qquad (3)$$

with beam energy $W_B$, the length of the space charge cloud l and the focal length f. The length of the cloud can be estimated by the distance between the two grounded electrodes (see fig. 1). Introducing the radial and longitudinal filling factors $\kappa_{r,l}$ [$0 \leq \kappa_{r,l} \leq 1$] and using Eq. (1), (2) and (3) the trapping efficiency can be expressed by:

$$\kappa_{r,l} = \frac{\overline{n}_e}{n_{e,r,l}} \qquad (4)$$

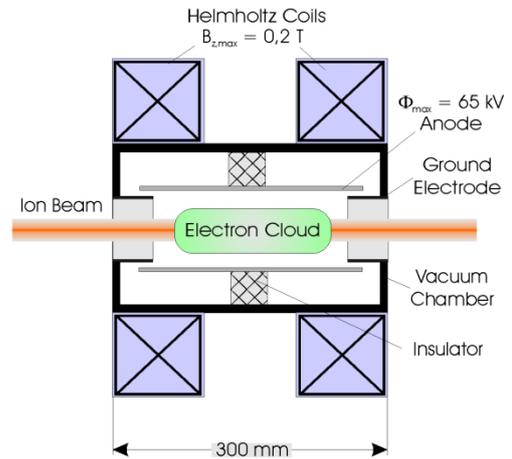

Figure 1: Cross-sectional view of a Gabor lens.

The density distribution of the electron cloud is influenced by collective effects. Numerical simulations and non-neutral plasma diagnostics help to characterise the dynamics of instabilities.

## NUMERICAL SIMULATIONS

  Two numerical codes are used for simulating the trapped electron column inside the Gabor lens.
  GABOR-M is a 2D hydrodynamics code, where the loss current is used as a free parameter. By recursive insertion of electrons and calculation of the force balance equation the lens volume is filled step by step. The result is an equilibrium density distribution based on the given parameters anode potential $V_A$, magnetic field $B_z$ and the loss current.

GAB_LENS is a 3D PIC (Particle-in-Cell) code using a constant number of macroparticles. The electrons are homogeneously distributed over the entire lens volume as a starting condition. The particle motion inside the lens is followed by a symplectic middle-step algorithm and every lost particle is immediately re-generated inside the lens, with random position and random momentum direction. Consequently, the number of particles is conserved but overall, energy is not. The macrocanonical state is approaching the most probable equilibrium state.

GABOR-M and GAB_LENS were developed to simulate the dynamic behaviour of stable confined electron columns, when electron losses and production rates are in balance. The work function of a Gabor lens depends on transverse and longitudinal confinement.

Charged particles are moving transversally on cycloidal paths in presence of the magnetic field with an average radius depending on the E over B ratio. This radius defines the maximum possible transverse extent of the entire trapped column due to the losses on the surface of the positive anode. Longitudinally, the average lifetime of the escaped electrons also depends on the E over B ratio and is important for effective field calculation.

Self-organized collective dynamics can be observed in simulations and were confirmed by measurements. Some examples are showed in Fig.2.

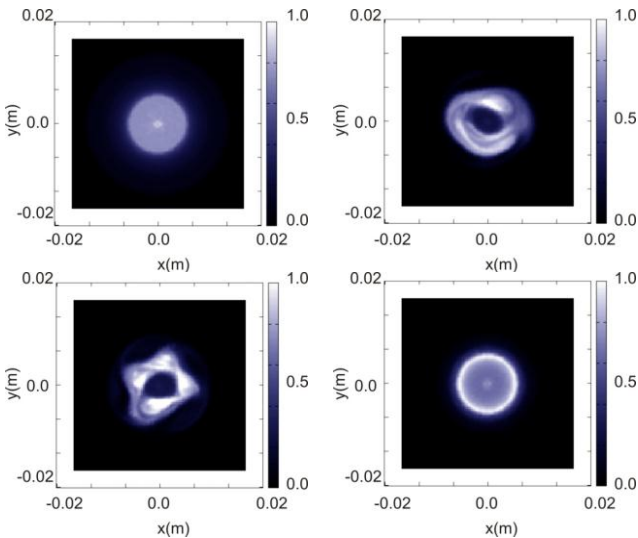

Figure 2: Comparison of transverse equilibrium density distributions for different conditions. The typical hollow electron distribution can be established in an overfilled Gabor lens. As a consequence the Diocotron azimuthal instability rises. Colour code: normalized density.

## DIAGNOSTIC

Different types of space charge lenses have been designed, built and tested in small-scale table top experiments.

They all provide several features to investigate different aspects of the confined electron cloud, e.g. electron density and temperature, electron production rates, and electron cloud dynamics.

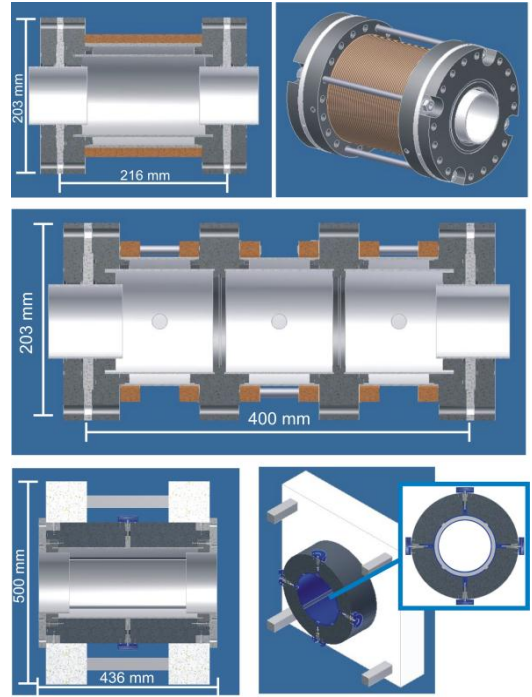

Figure 3: Different designs of space charge lenses (Top, Middle, Bottom).

Figure 3 (top) shows the design of the Gabor lens for beam energies up to 60keV [3]. The focussing capabilities have been studied in detail by [6]. The other pictured designs are scaled versions of the original one.
In the following the principles of the non-neutral plasma diagnostics used are explained. Selected results are discussed for the different lens types.

### Electron Production

The electrons are considered to be created by electron impact ionisation of the residual gas atoms. Because of the positive potential of the anode the produced ions are extracted. Figure 4 shows the ignition curves for the system extended in z-direction in comparison to the original design. In this context the ignition of the non-neutral plasma means that the electron production dominates the losses.

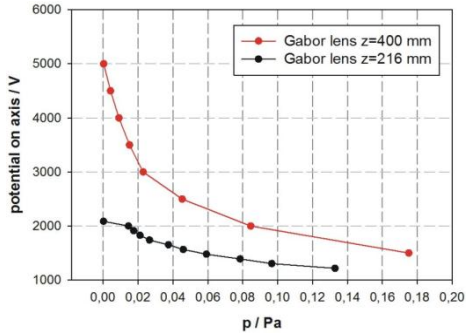

Figure 4: Ignition curves for different lengths of space charge lenses.

Because of the higher probability of electrons production at the edge of the potential well the assumed mean kinetic energy of the electron ensemble is much higher.

However, the short filling times cannot be explained by the sole interaction of the ions and electrons with residual gas. Therefore, the impact of particle losses on the production rate has to be considered as well.

Latest results suggest electron losses on the anode as further production mechanisms. As a result x-ray radiation can be detected.

*Electron Density Measurement*

The localisation of the production regions allows the determination of the mean electron density.

The maximum ion energy results from the anode potential. Because of the electron confinement its potential is reduced. The difference of the maximum possible energy to the actual measured energy enables the calculation of the mean electron density by the use of the theoretical confinement condition [4]:

$$n_e = \frac{4\varepsilon_0 \Delta\Phi}{er^2}. \quad (5)$$

The energy of the extracted ions can be measured using a momentum spectrometer. The two detected peaks are related to different production regions within the lens volume. This result is confirmed by comparing the measured ion energy to the numerical simulation (see Fig. 5).

Peak 1 is a result of the production of ions in a small region near the anode, which is mainly caused by electron losses. Peak 2 provides information about the electron density in the centre of the lens.

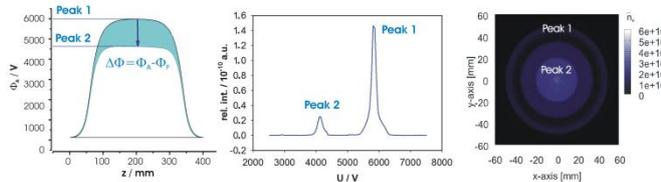

Figure 5: Measurement scheme of the mean electron density.

In the Gabor space charge lens mean electron densities of about $10^{14}$ m$^{-3}$ can be achieved.

*Electron Temperature Measurement*

Conventional non-interceptive diagnostic methods for the temperature determination are not applicable to non-neutral plasmas. This is due to the absence of important recombination processes. Additionally, the electron densities are too low to assume the plasma to be in a partially local thermal equilibrium.

It is currently under investigation whether the electron temperature can be determined by the measurement of the intensity ratio of two excited states in helium and comparing it to the ratio of their optical emission cross sections.

Therefore, an experiment was set up to measure optical emission cross sections for helium. The measurement scheme is shown in Figure 6.

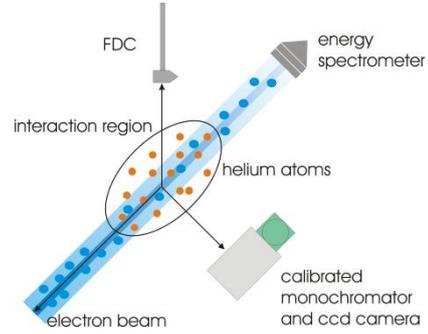

Figure 6: Measurement scheme of optical emission cross section. Electron current was measured by the use of a Faraday cup (FDC).

First comparisons of measured spectra (see Fig. 7) show a difference in radiative transitions for electron impact excitation between the case of an electron beam and the case of the confined electron cloud, especially for a wavelength of 588nm.

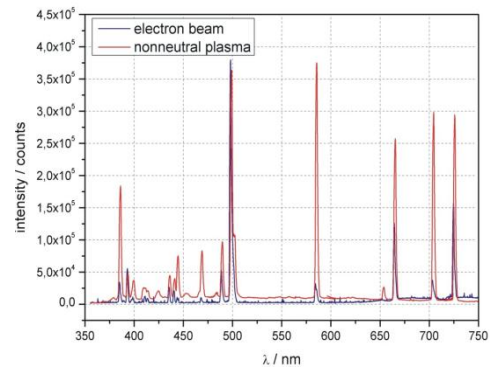

Figure 7: Comparison of measured spectra.

It still needs to be studied if this is due to multiple excitations or the difference in the electron energy distribution function.

*Electron Cloud Dynamics*

The evolution of plasma instabilities mainly generated by the variation of the local density and energy distribution is studied by time-resolved diagnostics [5].

In this case the oscillation of the measured ion current provides the trigger signal for the optical exposure. The change in the symmetry of the observed light density distribution is detected for different time steps along the path of the measured ion current signal (see Fig. 8).

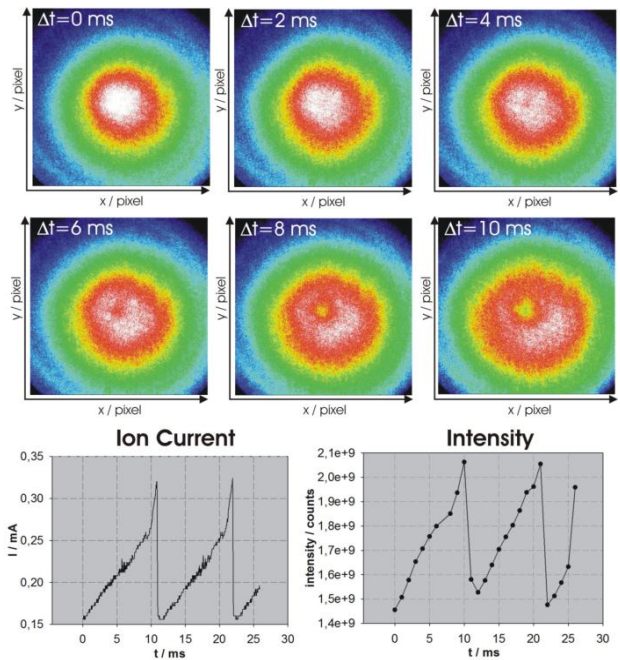

Figure 8: Time-resolved measurement of a non-neutral plasma instability (top), related ion current and intensity of the light density distribution (bottom).

The created hollow profile might also be the origin of the well-known Diocotron instability. The PIC simulation using GAB_LENS verifies the collective behaviour because of the confining fields. However, the interaction of electrons with residual gas atoms as well as the lifetime of ions in the system has to be considered, too, when describing the electron cloud dynamics.

## EXPERIMENTS USING ION BEAMS

The focussing capability was proven by the use of a $He^+$ beam with an energy of 440keV [6]. In a first step the average electron density and the trapping efficiency were estimated from the focusing effect on the ion beam using Eqs. 3 and 4. Figure 9 (left) shows the trapping efficiency as a function of the magnetic and electrostatic confinement. Following the work function of the Gabor lens $\kappa$ reaches values of about 50%. Increasing the magnetic field and the anode voltage along the optimum work function leads to a decrease of the trapping efficiency. This observation might be explained by the rise of the electron temperature.

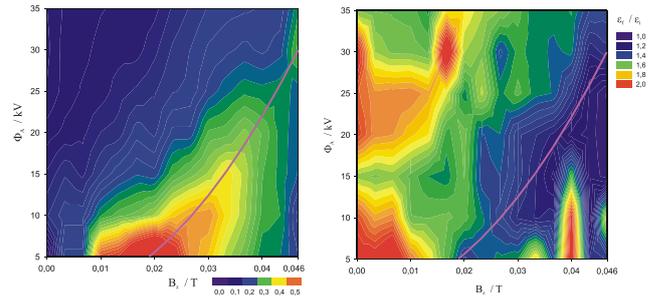

Figure 9: Measured trapping efficiency $\kappa$ (left) and mapping quality expressed by the emittance growth (right) as a function of transverse and longitudinal confinement, beam properties: $He^+$, $W_b$=440keV, I=1.2mA.

Emittance measurements were performed to estimate the mapping quality of the Gabor lens. The emittance growth along the work function as shown in Figure 9 (right) is negligible. A large decline of the beam quality occurs in the regime of induced plasma instabilities. The knowledge of the different plasma states within a Gabor lens would open this lens type to accelerator physics applications as a superior focussing device.

## CONCLUSIONS

Gabor lenses are under investigation as a powerful focussing device for intense heavy ion beams. This lens type enables the possibility to study the behaviour of stable confined electron clouds. Therefore, diagnostic tools were developed to measure non-neutral plasma properties as a function of the confinement condition. Numerical models are used to verify the experimental results and to study the collective phenomena in more detail. Furthermore, the interaction of an ion beam with the confined electron cloud can be studied.